# Lateral and axial resolution criteria in incoherent and coherent optics and holography, near- and far-field regimes

Tatiana Latychevskaia

ABSTRACT

This work presents an overview of the spatial resolution criteria in classical optics, digital optics and holography. Although the classical Abbe and Rayleigh resolution criteria have been thoroughly discussed in the literature, there are still several issues which still need to be addressed, for example the axial resolution criterion for coherent and incoherent radiation, which is a crucial parameter of three-dimensional (3D) imaging, the resolution criteria in the Fresnel regime, and the lateral and axial resolution criteria in digital optics and holography. This work discusses these issues and provides a simple guide for which resolution criteria should be applied in each particular imaging scheme: coherent/incoherent, far- and near-field, lateral and axial resolution. Different resolution criteria such as two-point resolution and the resolution obtained from the image spectrum (diffraction pattern) are compared and demonstrated with simulated examples. Resolution criteria for spatial lateral and axial resolution are derived, and their application in imaging with coherent and incoherent (noncoherent) waves is considered. It is shown that for coherent light, the classical Abbe and Rayleigh resolution criteria do not provide an accurate estimation of the lateral and axial resolution. Lateral and axial resolution criteria based on an evaluation of the spectrum of the diffracted wave provide a more precise estimation of the resolution for coherent and incoherent light. It is also shown that resolution criteria derived in approximation of the far-field imaging regime can be applied for the near-field (Fresnel) regime.

# 1. Introduction

Modern optics allows three-dimensional (3D) imaging at high resolution, where the resolution is an important parameter of the performance of the optical imaging system. The resolution is typically evaluated using the Abbe criterion, which firstly only evaluates the lateral resolution, and secondly does not provide correct results for coherent radiation. Moreover, the Abbe and the Rayleigh resolution criteria were derived assuming a far-field diffraction regime, and the question on how to evaluate resolution in the Fresnel diffraction regime remains open. The situation is even worse for axial resolution, since in most cases only the lateral resolution is considered, the axial resolution is often not addressed, although this may be a crucial parameter for the 3D imaging properties of the optical system. Also, in modern imaging techniques, such as coherent diffractive imaging where the resolution criteria are adapted from crystallography, other resolution criteria are applied. Below we discuss these issues and provide a simple guide for which resolution criteria should be applied in a particular imaging scheme. An overview of techniques for improving the resolution is beyond the scope of this manuscript, and a good review of these methods is provided by den Dekker and van den Bos [1].

## 2. Lateral and axial resolution

Before we quantitatively address the resolution criteria, we introduce two types of resolution, lateral and axial resolution, both of which are important in the characterization of an optical system. An optical system forms a 3D image of an object by re-focusing the wavefront scattered by the object. When a camera or a screen is placed in the in-focus plane, a 2D image of the sample is obtained. The quality of this 2D image is characterized by the lateral resolution $R_{\text{Lateral}}$, which is the resolution in the image plane $(x, y)$. Another important parameter of an optical system is the axial resolution $R_{\text{Axial}}$, which is the resolution of the 3D image along the optical axis. Typically, lateral resolution is better than axial resolution, since lateral resolution is inversely proportional to the numerical aperture of the system NA, while the axial resolution is inversely proportional to the squared numerical aperture $NA^2$ (as derived below). In most cases, only lateral resolution is mentioned and this is simply referred to as "resolution". In this study, both the lateral and the axial resolutions are considered.

## 3. Classical optics resolution criteria

This section is organized in chronological order. It should be noted that at the time when Abbe and Rayleigh derived their basic resolution criteria, there was no coherent light, and these formulae therefore describe resolution obtained with incoherent light.

### 3.1 Lateral resolution

### 3.1. Airy pattern

In 1835, Airy reported that light diffracted on a circular aperture exhibits concentric rings of alternating intensity maxima and minima - Airy patterns [2]. In his article, Airy tabulated the values of the function describing the intensity distribution, but he did not mention that they can be expressed through Bessel functions [3].

The resolution criteria can be derived by considering how well an optical system can image an infinitesimal point source. A point source is imaged using a $4f$ optical system, where the distance between the source and the lens is $4f$ and the distance between the lens of radius $a$ and the detector (image plane) is $z = 2f$. The distribution of the wavefront at the detector (image plane) is given by a Bessel function (see Appendix A):

$$U(x, y) = 2U_0 \frac{J_1\left(\frac{2\pi a q}{\lambda z}\right)}{\left(\frac{2\pi a q}{\lambda z}\right)}$$

and the intensity distribution is given by:

$$I(x, y) = I_0 \left|\frac{2J_1\left(\frac{2\pi a q}{\lambda z}\right)}{\left(\frac{2\pi a q}{\lambda z}\right)}\right|^2 \qquad (1)$$

where $\lambda$ is the wavelength, $J_1(...)$ is a Bessel function of the first kind, $q = \sqrt{x^2 + y^2}$ is the coordinate in the image plane, and $I_0 = |U_0|^2$ is the intensity at the center of the diffraction pattern.

### 3.1.2 Abbe resolution criterion

In 1873 - 1876, Abbe was developing optical microscopes at Zeiss, and in 1873 he published a paper on the resolution limit of an optical microscope. In this paper, Abbe did not provide any mathematical equations, and simply stated that the smallest object resolvable by a microscope cannot be smaller than half a wavelength [4], page 456. It was only in 1882 that he published a paper in which the lateral resolution limit was provided in form of an equation [5]:

$$R_{\text{Lateral}}^{\text{Abbe}} = \frac{\lambda}{2n \sin \vartheta_{\max}} = \frac{\lambda}{2\text{NA}}, \qquad (2)$$

where $\sin \vartheta_{\max}$ is the maximal scattering angle detected by the optical system, $n$ is the refractive index and $\text{NA} = n \sin \vartheta_{\max}$ is the numerical aperture of the system. For simplicity, we will assume $n = 1$ in the following. $\text{NA} \approx a/z$, where $z$ is the distance from the aperture, and $a$ is the radius of the aperture.

### 3.1.3 Two-point resolution and the Rayleigh criteria

One resolution criteria can be formulated based on the observation of how well images of two point sources are resolved. This approach is conventionally known as two-point resolution, and is illustrated in Fig. 1.

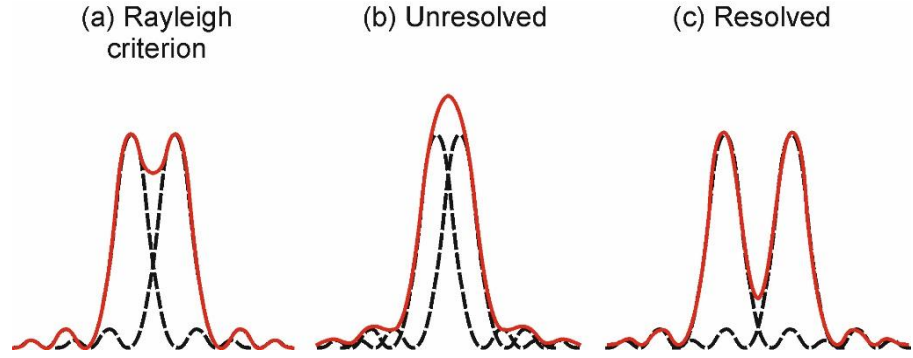

Fig. 1. Images of two point sources and resolution criteria. (a) Images of two point sources that are just resolved, according to the Rayleigh criterion. (b) Images of two point sources that are not resolved. (c) Images of two point sources that are resolved.

In 1879, Rayleigh formulated the following resolution criterion [6]: two point sources are regarded as *just resolved* when the zero-order diffraction maximum of one diffraction pattern coincides with the first minimum of the other, as illustrated in Fig. 1(a).

The first zero of the Bessel function occurs when its argument equals 3.83. By applying the Rayleigh criterion to the intensity distribution of an image of a point source (Eq. 1), we obtain the condition $\frac{2\pi a q}{\lambda z} = 3.83,$ which gives

$$R_{\text{Lateral}}^{\text{Rayleigh}} = 0.61 \frac{\lambda z}{a} = 0.61 \frac{\lambda}{\text{NA}}. \tag{3}$$

When the distance between the zero-order diffraction maxima of the two diffraction patterns is less than the distance given by the Rayleigh criterion, the two point sources are not resolved, as illustrated in Fig. 1(b).

Two point sources are *resolved* when the zero-order diffraction maximum of one diffraction pattern coincides with the first-order diffraction maximum of the other, as shown in Fig. 1(c). The first maximum of the intensity given by Eq. 1 occurs when $\frac{2\pi a q}{\lambda z} = 5.14,$ which gives

$$R_{\text{Lateral}}^{\text{Resolved}} = 0.82 \frac{\lambda z}{a} = 0.82 \frac{\lambda}{\text{NA}}.$$

A detailed historical overview of how these optical resolution criteria were formulated in the original manuscripts and re-formulated into their modern form is provided in a book by de Villiers and Pike [7].

## 3.2. Axial resolution criterion

When a point source is imaged by an optical system, the axial distribution of the focused wavefront is described by the sinc function (see Appendix B):

$$U(0,0,\Delta z) = U_0 \exp\left(-\frac{i\beta a^2}{2}\right) \frac{\sin\left(\frac{\beta a^2}{2}\right)}{\frac{\beta a^2}{2}},$$

where $\beta = \frac{\pi \Delta z}{\lambda z^2}$, and $\Delta z$ is the defocus distance from the in-focus position at $z$. The intensity distribution is given by

$$I(0,0,\Delta z) = I_0 \left| \frac{\sin\left(\frac{\beta a^2}{2}\right)}{\frac{\beta a^2}{2}} \right|^2.$$

The axial resolution can be defined using the Rayleigh criterion: two point sources are regarded as just resolved when the zero-order diffraction maximum of one diffraction pattern coincides with the first minimum of the other. The first minimum occurs at:

$$\frac{\beta a^2}{2} = \pi,$$

which corresponds to the distance $\Delta z'$ that provides the axial resolution criterion

$$R_{\text{Axial}}^{\text{Rayleigh}} = \Delta z' = \frac{2\lambda z^2}{a^2} = \frac{2\lambda}{\text{NA}^2}, \qquad (4)$$

where $\text{NA} \approx \frac{a}{z}$ is the numerical aperture of the optical system as defined above.

## 4. Diffraction pattern resolution criteria

Diffraction pattern resolution is determined by the highest detectable frequency in the image spectrum (diffraction pattern). This criterion is often applied in X-ray or electron crystallography and the coherent diffraction imaging of non-periodic samples [8] where a diffraction pattern is acquired. To derive the lateral and the axial resolution criteria in this case, we first provide several formulae describing the formation of a diffraction pattern.

When a plane wave is incident on an object $o(\vec{r}_0)$, where $\vec{r}_0 = (x_0, y_0, z_0)$ is the coordinate in the object domain, the distribution of the scattered wavefront in the far-field regime is given by

$$u(\vec{r}) = \iiint \exp(ikz_0) o(\vec{r}_0) \frac{\exp(ik|\vec{r}-\vec{r}_0|)}{|\vec{r}-\vec{r}_0|} d\vec{r}_0 \quad (5)$$

where $\vec{r} = (x, y, z)$ is the coordinate in the far-field domain. The argument of the second exponent in the integral can be expanded as follows:

$$|\vec{r}-\vec{r}_0| = \sqrt{r_0^2 - 2\vec{r}\vec{r}_0 + r^2} \approx r - \frac{\vec{r}\vec{r}_0}{r}. \quad (6)$$

Next, the scattering vector is introduced as follows:

$$\vec{K} = k\frac{\vec{r}}{r} = k\left(\frac{x}{r}, \frac{y}{r}, \frac{z}{r}\right) = (K_x, K_y, K_z), \quad (7)$$

where $K_x^2 + K_y^2 + K_z^2 = k^2$. By substituting Eqs. 6 and 7 into Eq. 5 we obtain:

$$u(\vec{K}) \propto \iiint \exp(ikz_0) o(\vec{r}_0) \exp(-i\vec{K}\vec{r}_0) d\vec{r}_0 =$$

$$= \iiint \exp(ikz_0) o(x_0, y_0, z_0) \exp\left[-i(K_x x_0 + K_y y_0)\right] \exp\left[-iz_0 \sqrt{k^2 - K_x^2 - K_y^2}\right] dx_0 dy_0,$$

where the resulting far-field distribution is in $(K_x, K_y)$ coordinates.

## 4.1. Lateral resolution criterion

A wavefront diffracted on a periodical object with period $d$ will create a peak in the far-field intensity distribution (diffraction pattern) at

$$K_{x,y}^{(d)} = \frac{2\pi}{d}.$$

On the other hand, the largest detected value of $K_{x,y}$ is acquired at the largest diffraction angle $\sin\vartheta_{\max}$:

$$K_{x,y,\max} = \frac{2\pi}{\lambda}\sin\vartheta_{\max}.$$

Thus, the smallest detectable period that defines the lateral resolution is given by:

$$R_{\text{Lateral}}^{(DP)} = d_{x,y} = \frac{2\pi}{K_{x,y}^{(d)}} = \frac{\lambda}{\sin\vartheta_{\max}} = \frac{\lambda}{\text{NA}}. \quad (8)$$

The obtained lateral resolution is twice as large as that obtained with the Abbe resolution criterion.

## 4.2. Axial resolution criterion

The axial resolution is given by the available spread of $K_z$ values:

$$R_{\text{Axial}}^{(DP)} = \frac{2\pi}{\Delta K_z},$$

where $\Delta K_z$ is the spread of the $K_z$ values. In the center of the detector $K_z = k = \frac{2\pi}{\lambda}$, and at the rim of the detector, $K_z = \sqrt{k^2 - K_{x,y,\max}^2}$ which gives $\Delta K_z$

$$\Delta K_z = k - \sqrt{k^2 - K_{x,y,\max}^2} \approx \frac{K_{x,y,\max}^2}{2k} = \frac{2\pi}{\lambda} \frac{\sin^2 \vartheta_{\max}}{2},$$

and thus

$$R_{\text{Axial}}^{(\text{DP})} = \frac{2\pi}{\Delta K_z} = \frac{2\lambda}{\sin^2 \vartheta_{\max}} = \frac{2\lambda}{\text{NA}^2}. \tag{9}$$

The axial resolution criterion obtained in this way is the same as the axial resolution criterion obtained by considering the imaging of a point source in an optical system.

## 5. Effect of coherence

### 5.1. Lateral resolution

Examples of the effect of coherence on lateral resolution can be found in optical textbooks, for example [9, 10]. For coherent light, the Rayleigh resolution criterion given in Eq. 3 becomes [9]:

$$R_{\text{Lateral}}^{(\text{Rayleigh,coherent})} = 0.82 \frac{\lambda}{\text{NA}}. \tag{10}$$

Numerical examples of two-point resolution with coherent and incoherent light are shown in Fig. 2. The simulation procedure is described in Appendix C.

For incoherent light, the Rayleigh resolution criterion provides the most accurate lateral resolution distance: the two points are just resolved when the distance between them is equal to the Rayleigh criterion distance, as shown in Fig. 2(e) and (f). For coherent light, when the distance between the two points is equal to the Abbe resolution or the Rayleigh resolution distance, the two points appear unresolved. For coherent light, the Rayleigh resolution criterion for coherent light (Eq. 10) provides a more accurate result, as shown in Fig. 2(g) and (i). For both coherent and incoherent light, when the distance between two sources is the same as the distance given by the diffraction resolution criterion in Eq. 8, the two points are well resolved, as shown in Fig. 2(j), (k) and (l). It should be noted that when the images of the two point sources are not resolved, they appear in the wrong lateral positions, for both coherent and incoherent light. This effect for the lateral resolution has been previously discussed by Grimes and Thompson [11] and Goldstein [12].

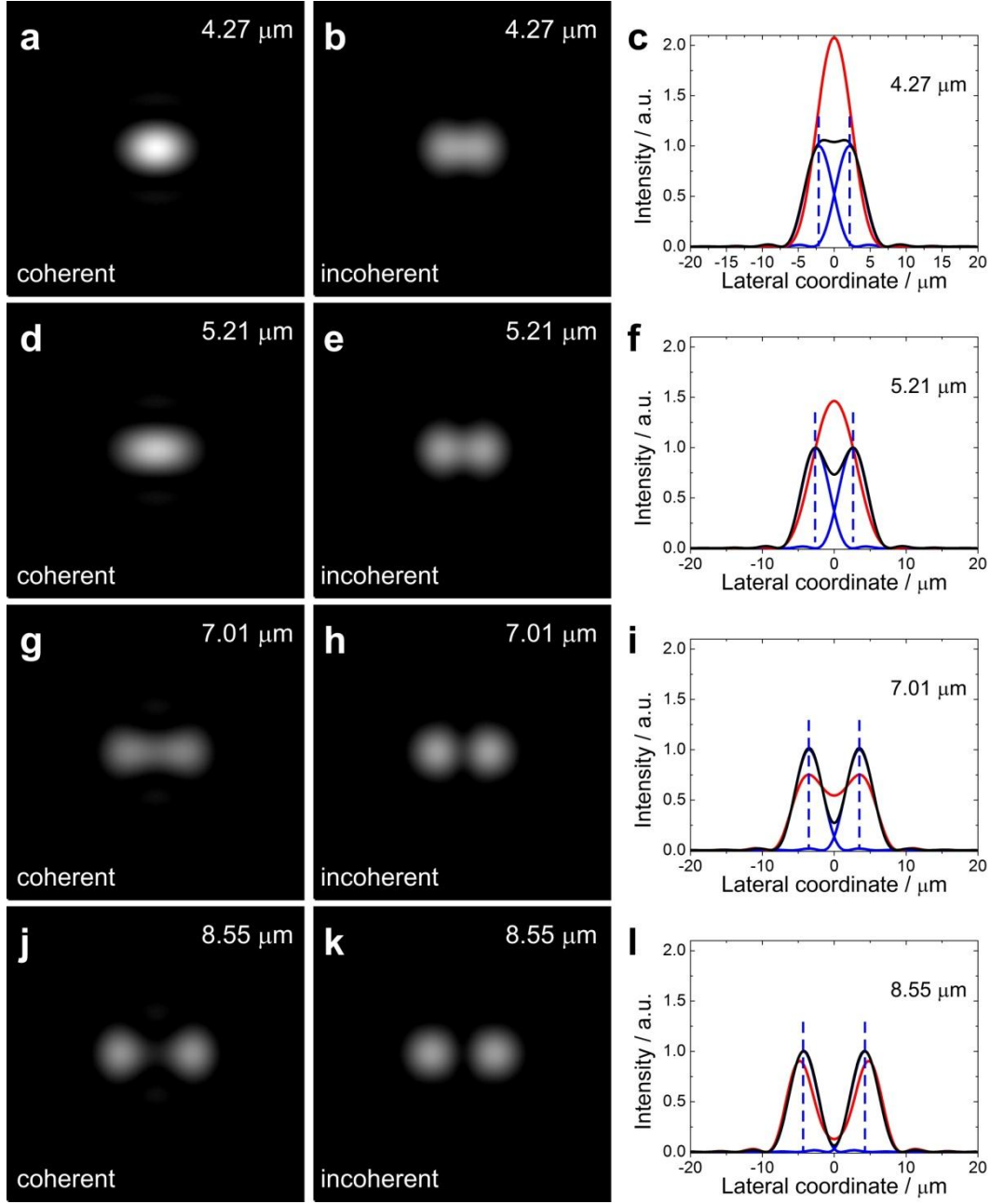

Fig. 2. Effect of coherence on lateral resolution. The parameters of the imaging system are: wavelength $\lambda = 532$ nm, aperture (lens) diameter 25 mm, aperture (lens)-to-image plane distance $z = 0.2$ m. With these parameters, the estimated lateral resolution of the imaging system, according to four different criteria, is as follows: (1) Abbe criterion (Eq. 2): $R_{\text{Lateral}}^{(\text{Abbe})} = 4.27$ μm, (2) Rayleigh criterion (Eq. 3): $R_{\text{Lateral}}^{(\text{Rayleigh})} = 5.21$ μm, (3) Rayleigh criterion for coherent light (Eq. 10): $R_{\text{Lateral}}^{(\text{Rayleigh,coherent})} = 7.01$ μm and (4) diffraction pattern resolution criterion (Eq. 8): $R_{\text{Lateral}}^{(\text{DP})} = 8.55$ μm. Two point sources are separated by distances of (a) – (c) 4.27 μm, (d) – (f) 5.21 μm, (g) - (i) 7.01 μm and (j) – (l) 8.55 μm. The intensity of the 2D images and in the plots is normalized so that the maximal intensity of one point source image is 1 a.u. The intensity values in each 2D image range between 0 and 2.1 a.u. The red curves in (c), (f), (i) and (l) are the intensity profiles of two coherent sources. The black curves in (c), (f), (i) and (l) are the intensity profiles of two incoherent sources. The blue curves in (c), (f), (i) and (l) are the intensity profiles of the individually imaged sources. The blue dashed lines in (c), (f), (i) and (l) indicate the correct positions of the individually imaged sources.

## 5.2. Axial resolution

The effect of coherence on axial resolution is illustrated in Fig. 3. The simulation procedure is described in Appendix C.

For both coherent and incoherent light, the two sources appear to be resolved when the distance between the sources satisfies the axial resolution criterion, as given by Eqs. 4 and 9. It should be noted that in the case of coherent light, despite being clearly resolved one from another, the images of the sources appear at incorrect axial positions.

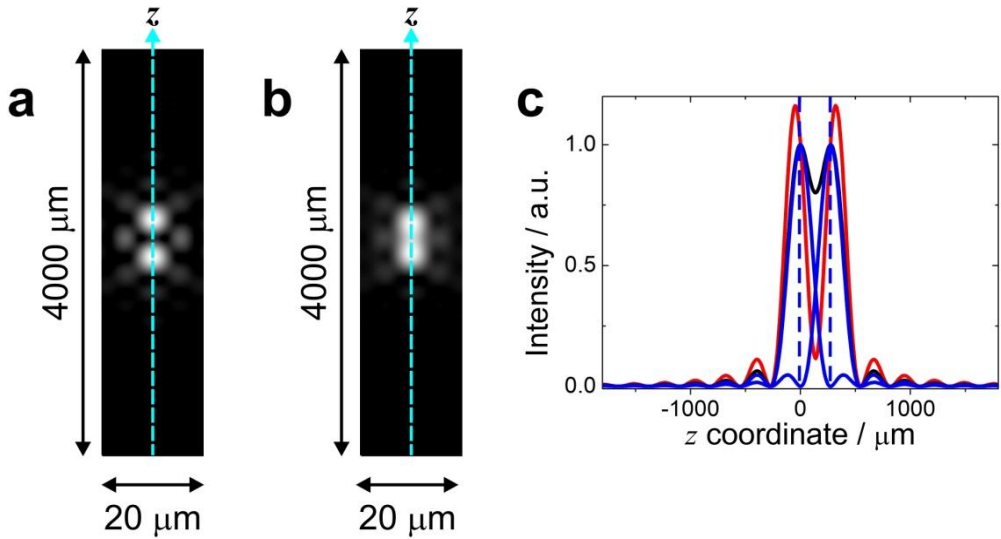

Fig. 3 Effect of coherence on axial resolution. The parameters of the imaging system are: wavelength $\lambda = 532$ nm, aperture (lens) diameter 25 mm, aperture (lens)-to-image plane distance $z = 0.2$ m. With these parameters, the estimated axial resolution of the imaging system is (Eqs. 4 and 9): $R_{\text{Axial}}^{(\text{Rayleigh})} = R_{\text{Axial}}^{(\text{DP})} = 275$ μm. The two point sources are separated by 275 μm. (a) and (b) show the 2D intensity plots when the two sources are (a) coherently and (b) incoherently imaged. The intensity in the 2D images and in the plot is normalized so that the maximal intensity of one point source image is 1 a.u. (c)

The intensity profiles along the $z$-axis. The red curves are the intensity profiles of two coherent sources. The black curves are the intensity profiles of two incoherent sources. The blue curves are the intensity profiles of the individually imaged sources. The blue dashed lines indicate the correct positions of the individually imaged sources.

# 6. Resolution criteria for near- and far-field imaging

The resolution criteria discussed above are derived assuming the far-field (Fraunhofer) diffraction regime. For example, Airy patterns are only obtained as a result of far-field diffraction on a circular aperture. However, there is often a need to evaluate the resolution of an optical system when the imaging is done not in the far- but in the near-field imaging regime. As "near-field imaging regime" we call the imaging at a $z$-distance which is shorter than that satisfying the far-field regime. It should not be confused with near-field optics operating in the sub-wavelength distance regime [13]. Diffraction in the Fresnel (near-field) regime is typically utilized in optics and holography, for example in in-line holography with plane waves [14, 15], focal series [16] etc.

The difference between the near- and far-field regimes is best explained by the imaging integral. A scattered wave in the Fresnel approximation is written as

$$u(x,y) = \iint o(x_0, y_0) \exp(ik|\vec{r}-\vec{r}_0|) dx_0 dy_0 \approx$$
$$\approx \iint o(x_0, y_0) \exp\left\{\frac{i\pi}{\lambda z}\left[(x-x_0)^2 + (y-y_0)^2\right]\right\} dx_0 dy_0, \quad (11)$$

where the approximation was obtained by a Taylor series expansion assuming that [10]

$$z^3 \gg \frac{\pi}{4\lambda}\left[(x-x_0)^2 + (y-y_0)^2\right]^2_{max},$$

which defines the Fresnel diffraction regime. A much stronger approximation:

$$z^2 \gg \frac{\pi}{\lambda}\left(x_0^2 + y_0^2\right)_{max},$$

at which the integral in Eq. 11 turns into a 2D Fourier transform:

$$u(x,y) \approx \iint o(x_0, y_0) \exp\left\{\frac{i\pi}{\lambda z}\left[(x-x_0)^2 + (y-y_0)^2\right]\right\} dx_0 dy_0 \approx$$
$$\approx \iint o(x_0, y_0) \exp\left[-\frac{i2\pi}{\lambda z}(xx_0 + yy_0)\right] dx_0 dy_0,$$

defines the Fraunhofer diffraction regime. An optical system can be assigned a Fresnel number:

$$F = \frac{a^2}{z\lambda},$$

where $a$ is the radius of the aperture and $z$ is the distance from the aperture. At $F \ll 1$ the diffraction is Fraunhofer diffraction, while at $F \geq 1$ the diffraction is Fresnel diffraction.

The correct approximation for propagation in the near field is provided by the angular spectrum method (ASM) [10, 17]. In the ASM, the diffracted wavefront is calculated as:

$$u(x, y) = \text{FT}^{-1}\left\{\text{FT}\left[o(x_0, y_0)\right]\exp\left(\frac{2\pi i z}{\lambda}\sqrt{1-\alpha^2-\beta^2}\right)\right\},$$

where

$$\text{FT}\left[o(x_0, y_0)\right] = \iint o(x_0, y_0)\exp\left[-2\pi i\left(\frac{\alpha}{\lambda}x_0 + \frac{\beta}{\lambda}y_0\right)\right]dx_0 dy_0.$$

Propagation in the ASM thus includes three steps: (1) a Fourier transform of $o(x_0, y_0)$, (2) multiplication with the factor $\exp\left(\frac{2\pi i z}{\lambda}\sqrt{1-\alpha^2-\beta^2}\right)$, and (3) inverse Fourier transform. The last step of the ASM can be considered as propagation from "the far-field" to the image plane. We can therefore apply the resolution criteria derived in the far-field approximation if we consider the distribution $\text{FT}\left[o(x_0, y_0)\right]\exp\left(\frac{2\pi i z}{\lambda}\sqrt{1-\alpha^2-\beta^2}\right)$ as the "far-field" wavefront distribution. Here, $\alpha$ and $\beta$ are directional cosines that define the scattering wave vector as $\vec{k} = \frac{2\pi}{\lambda}\left(\alpha\vec{e}_x + \beta\vec{e}_y + \gamma\vec{e}_z\right)$. By comparing this definition to Eq. 7, we obtain: $\alpha = K_x$ and $\beta = K_y$. A similar discussion to that in Section 4 can be applied and the near-field (NF) resolution criteria are obtained:

$$R_{\text{Lateral}}^{(\text{NF})} = \frac{\lambda}{\text{NA}}$$

and

$$R_{\text{Axial}}^{(\text{NF})} = \frac{2\lambda}{\text{NA}^2}.$$

# 7. Resolution in digital optics and holography

In digital optics and holography, wavefronts are sampled with a finite number of pixels. Therefore, in addition to the other parameters of the optical system, the number of pixels and the pixel size can affect the resulting resolution.

## 7.1. Lateral resolution

In digital holography, when a wavefront is numerically propagated from the measured intensity (hologram) plane to the object plane, the pixel size in the object plane is given by [18]:

$$\Delta_0 = \frac{\lambda z}{N\Delta} \approx \frac{\lambda}{2\mathrm{NA}}, \tag{12}$$

where $\Delta$ is the pixel size in the measured intensity distribution (hologram), $N$ is the number of pixels, $z$ is the distance between the two planes, the screen size is given by $N\Delta = s$, and the numerical aperture of the system is $\mathrm{NA} \approx \frac{s}{2z}$. Equation 12 provides the pixel size in the object plane. However to resolve two points, at least two pixels are needed, leading to

$$R_{\mathrm{Lateral}}^{(\mathrm{DH})} = 2\Delta_0 \approx \frac{\lambda}{\mathrm{NA}},$$

which represents the lateral resolution criterion in digital optics and holography (DH), where we have substituted $\Delta_0$ from Eq. 12. The same result was previously obtained for the lateral resolution criterion in digital holography [19], where the effects of sampling were also investigated.

## 7.2 Axial resolution

When a wavefront is numerically propagated from the measured intensity (hologram) plane to the object plane, the $z$-position of the reconstruction plane can be arbitrary selected, and thus there is no limit on how finely the reconstructed wavefront can be sampled along the $z$-axis. The axial resolution limit is thus the same as for light optics or diffraction:

$$R_{\mathrm{Axial}}^{(\mathrm{DH})} = \frac{2\lambda}{\mathrm{NA}^2},$$

which for a digitally sampled intensity distribution becomes:

$$R_{\mathrm{Axial}}^{(\mathrm{DH})} = \frac{8\lambda z^2}{N^2 \Delta^2}.$$

## 8. Discussion and conclusions

In conclusion, we have shown that for incoherent light, the Abbe and Rayleigh criteria provide accurate estimations of the lateral resolution. However, these criteria do not provide an accurate estimation of the lateral resolution for coherent light. The Rayleigh criterion can be re-formulated for coherent light. Resolution criteria based on an evaluation of the spectrum of the diffracted wave provide an accurate estimation of the lateral and axial resolution for both coherent and incoherent light. In summary, these resolution criteria are as follows:

Lateral resolution criterion for incoherent light

$$R_{\text{Lateral}}^{(\text{Rayleigh})} = 0.61 \frac{\lambda}{\text{NA}}.$$

Lateral resolution criteria for coherent light

$$R_{\text{Lateral}}^{(\text{Rayleigh, coherent})} = 0.82 \frac{\lambda}{\text{NA}},$$

$$R_{\text{Lateral}}^{(\text{DP})} = \frac{\lambda}{\text{NA}}.$$

Axial resolution criterion for incoherent and coherent light

$$R_{\text{Axial}} = \frac{2\lambda}{\text{NA}^2}.$$

Lateral resolution criterion in digital optics and holography

$$R_{\text{Lateral}}^{(\text{DH})} = \frac{2\lambda z}{N\Delta}.$$

Axial resolution criterion in digital optics and holography

$$R_{\text{Axial}}^{(\text{DH})} = \frac{8\lambda z^2}{N^2 \Delta^2}.$$

Since in reality, sources of light, electrons, X-rays etc. exhibit partial coherence, it seems more reasonable to apply the resolution criteria based on evaluation of the spectrum of the diffracted wave. These resolution criteria can be also adapted for near-field imaging. Another important conclusion from the results presented here is that when point sources are not resolved, their apparent positions correspond only approximately to their actual positions.

# 9. Appendices

## 9.1 Appendix A: Wavefront reconstructed in the (x, y)-plane

A point source is imaged by a $4f$ optical system, where the distance between the source and the lens is $2f$ and the distance between the lens and the detector (image plane) is $z = 2f$. The distribution of the wavefront at the detector (plane) is given by:

$$U(x,y) = -\frac{i}{\lambda z}\exp(ikz)\exp\left[\frac{\pi i}{\lambda z}(x^2 + y^2)\right]\iint \exp\left[-\frac{2\pi i}{\lambda z}(xx_A + yy_A)\right]dx_A dy_A,$$

where integration is performed over the lens surface ("Aperture"). Since only the intensity is measured at the detector plane, the phase terms in front of the integral can be omitted. The integral can be solved in polar coordinates, and a detailed solution can be found in optical text books. The result is

$$U(x,y) = 2U_0 \frac{J\left(\frac{2\pi a q}{\lambda z}\right)}{\left(\frac{2\pi a q}{\lambda z}\right)},$$

where $U_0 = -\frac{\pi i a^2}{\lambda z}$.

## 9.2 Appendix B: Wavefront reconstructed along the z-axis

A point source positioned on an optical axis at $x_0 = y_0 = 0$ is imaged by a $4f$ optical system, where the distance between the source and the lens is $2f$ and the distance between the lens and the detector (image plane) is $z = 2f + \Delta z$. The distribution of the wavefront at the detector (image plane) is given by:

$$U(x,y,\Delta z) = -\frac{i}{\lambda(2f + \Delta z)} \times$$

$$\times \iint \exp\left[-\frac{\pi i \Delta z}{\lambda(2f)^2}(x_A^2 + y_A^2)\right]\exp\left[-\frac{2\pi i}{\lambda(2f + \Delta z)}(xx_A + yy_A)\right]dx_A dy_A,$$

where the integration is performed over the lens surface ("Aperture"). At $x = y = 0$ we obtain:

$$U(x,y,\Delta z) = -\frac{i}{2\lambda f}\iint \exp\left[-\frac{\pi i \Delta z}{\lambda(2f)^2}(x_A^2 + y_A^2)\right]dx_A dy_A.$$

We introduce the polar coordinates $x = \rho\cos\varphi$, $y = \rho\sin\varphi$ and rewrite this as

$$U(x,y,\Delta z) = -\frac{i}{2\lambda f}\int_0^a \int_0^{2\pi} \exp\left[-\frac{\pi i \Delta z}{\lambda(2f)^2}\rho^2\right]\rho d\rho d\varphi = -\frac{i\pi}{2\lambda f}\int_0^a \exp\left[-i\beta\rho^2\right]d\rho^2,$$

where $\beta = \dfrac{\pi \Delta z}{\lambda (2f)^2}$. The result of the integral is

$$U(x, y, \Delta z) \approx U_0 \exp\left(-\dfrac{i\beta a^2}{2}\right) \dfrac{\sin\left(\dfrac{\beta a^2}{2}\right)}{\left(\dfrac{\beta a^2}{2}\right)},$$

where $U_0 = -\dfrac{i\pi a^2}{2\lambda f}$.

## 9.3 Appendix C: Simulation procedure

The far-field distribution of a wavefront originating from a point source $i$ was simulated as a divergent spherical wave:

$$U_i(x', y') = \exp\left[\dfrac{2\pi i}{\lambda}\sqrt{(x'-x_i)^2 + y'^2 + (z'-z_i)^2}\right], \quad i = 1, 2 \tag{C1}$$

where $(x_i, 0, z_i)$, $i = 1, 2$ are the coordinates of the source. The wavefront in the image plane was calculated by employing the Fresnel-Huygens principle:

$$u_i(x, y, z) = \iint U_i(x', y') \exp\left[-\dfrac{2\pi i}{\lambda}\sqrt{(x'-x)^2 + (y'-y)^2 + (z'-z)^2}\right] dx' dy', \quad i = 1, 2. \tag{C2}$$

The integrals given by Eqs. C1 and C2 were programmed exactly as they stand, without approximations, and no fast Fourier transforms (FFT) were employed.

For incoherent waves, the intensity of the resulting image was calculated as:

$$I(x, y, z) = |u_1(x, y, z)|^2 + |u_2(x, y, z)|^2.$$

For coherent waves, the intensity of the resulting image was calculated as:

$$I(x, y, z) = |u_1(x, y, z) + u_2(x, y, z)|^2.$$